\documentclass{article}
\usepackage{arxiv}
\usepackage[utf8]{inputenc} 
\usepackage[T1]{fontenc}    
\usepackage{hyperref}       
\usepackage{url}  
\usepackage{algorithmic}
\usepackage{algorithm}
\usepackage{booktabs}       
\usepackage{amsfonts}  
\usepackage{amsmath}
\usepackage{nicefrac}       
\usepackage{microtype}     
\usepackage{lipsum}		
\usepackage{tikz,graphics,color,fullpage,float,epsf,caption,subcaption}
\usepackage[square, numbers, comma, sort&compress]{natbib} 
\usepackage{doi}

\title{MobileCaps: A Lightweight Model for Screening and Severity Analysis of COVID-19 Chest X-Ray Images}

\author{S J Pawan$^{a}$, Rahul Sankar$^{a}$, Amithash M Prabhudev $^{b}$, P A Mahesh $^{c}$,  K Prakashini$^{d}$, Sudha Kiran Das $^{e}$ and Jeny Rajan$^{a}$
\\
\\
$^{a}$Department of Computer Science and Engineering\\
National Institute of Technology Karnataka\\ 
Surathkal, India\\
\and
$^{b}$Department of Respiratory Medicine\\
Kasturba Medical College and Hospital, Manipal\\
Manipal Academy of Higher Education, Karnataka\\
\and
$^{c}$Department of Respiratory Medicine\\
J.S.S. Medical College, Mysore, India\\
JSS Academy of Higher Education and Research, India\\
\and
$^{d}$Department of Radiodiagnosis and Imaging\\
Kasturba Medical College and Hospital, Manipal\\
Manipal Academy of Higher Education, Karnataka, India\\
\and
$^{e}$Department of Radiology\\
J.S.S. Medical College, Mysore\\
JSS Academy of Higher Education and Research, India\\
}

\hypersetup{
pdftitle={MobileCaps: A Lightweight Model for Screening and Severity Analysis of COVID-19 Chest X-Ray Images},
pdfsubject={q-bio.NC, q-bio.QM},
pdfauthor={S J Pawan, Rahul Sankar, Amithash M Prabhudev, P A Mahesh, K Prakashini, Sudha Kiran Das and  Jeny Rajan },
pdfkeywords={Capsule Network, Convolutional Neural Network, Image Classification},
}

\begin{document}
\maketitle
\begin{abstract}
The world is going through a challenging phase due to the disastrous effect caused by the COVID-19 pandemic on the healthcare system and the economy. The rate of spreading, re-infections, post-COVID-19 symptoms, and the occurrence of new strands of COVID-19 have put the healthcare systems in disruption across the globe. Due to this, the task of accurately screening COVID-19 cases has become of utmost priority. Since the virus infects the respiratory system, Chest X-Ray (CXR) is an imaging modality that is adopted extensively for the initial screening. We have performed a comprehensive study that uses CXR images to identify COVID-19 cases and realized the necessity of having a more generalizable model. We utilize MobileNetV2 architecture as the feature extractor and integrate it into Capsule Networks to construct a fully automated, hybrid, and lightweight model termed as MobileCaps. MobileCaps is trained and evaluated on the publicly available COVIDx dataset with the model ensembling and Bayesian optimization strategies to efficiently classify CXR images of patients with COVID-19 from non-COVID-19 pneumonia and healthy cases. The proposed model is further evaluated on two additional RT-PCR confirmed datasets to demonstrate the generalizability. We also introduce MobileCaps-S and leverage it for performing severity assessment of CXR images of COVID-19 based on the Radiographic Assessment of Lung Edema (RALE) scoring technique. 
Our classification model achieved an overall recall of 91.60, 94.60, 92.20, and a precision of 98.50, 88.21, 92.62 for COVID-19, non-COVID-19 pneumonia, and healthy cases, respectively. Further, the severity assessment model attained an R$^2$ coefficient of 70.51. Owing to the fact that the proposed models have fewer trainable parameters than the state-of-the-art models reported in the literature, we believe our models will go a long way in aiding healthcare systems in the battle against the pandemic.
\end{abstract}


\section{Introduction}

The world is undergoing one of the greatest medical emergencies of all time. The outbreak of COVID-19 has perpetrated a massive blow on the health and economic conditions of millions across the globe. Viruses are said to be on the fence between living and non-living organisms. By themselves, they are not capable of doing anything, but once they enter into a living cell of a host, they can take control of the cell and replicate themselves, resulting in the damage of the host cell. Corona virus is an enveloped RNA virus, which is highly contagious \cite{who2020a}. The name \textit{corona} is originated from Latin, meaning \textit{crown} \cite{adhikari2020epidemiology}. The virus possesses crown-shaped spikes giving the appearance of the \textit{Solar Corona}. 
These viruses primarily infect birds and mammals, but their zoonotic nature can result in a cross-species infection leading to the formation of a dangerous variant infecting humans.
Some of the previous versions of \textit{coronaviruses} involved in human infection are Severe Acute Respiratory Syndrome-CoV of 2013 and Middle East Respiratory Syndrome-CoV of 2012 \cite{world2020critical}.
As per World Health Organization (WHO) guidelines, the terminology n-COVID-19 represents the novel Corona Virus Disease. 
Considering the outbreak affecting multiple countries in a repetitious fashion across the continent, WHO declared n-COVID-19 a pandemic on 11 March 2020. The predominant transmission mode of n-COVID-19 is through direct, indirect, or close contact with the infected people when their respiratory secretions are expelled as droplets when they cough, sneeze, sing or talk. Airborne transmission caused by the dissemination of droplet nuclei (aerosols) is evident in medical settings during aerosol-generating procedures and only hypothesized indoors with poor ventilation. Despite consistent evidence on the survival of SARS CoV2 on surfaces and contamination, the possibility of indirect transmission through fomites lacks specific reports. Other modes of transmission like faeco-oral, urine, serum, blood have not been proven yet \cite{world2020transmission}. The time from exposure to onset of symptoms may range from two to fourteen days. Common symptoms include cough, fever, shortness of breath, sore throat, fatigue, diarrhea, muscle pain, and abdominal pain. Most of the cases result in mild symptoms (above 80\%), while some progress to viral pneumonia and multi-organ failure. Old age people and people with pre-existing medical conditions (such as asthma, diabetes, and heart disease) are found more vulnerable to becoming severely ill \cite{world2020medical}. The high mutation rate and the asymptomatic nature of those infected inflate the drug/vaccine development complexity. Initially, the drugs such as chloroquine, hydroxychloroquine, remdesivir, dexamethasone were found effective in controlling the damages caused when the patient's immune system is affected, particularly those who need oxygen or ventilation support \cite{dong2020discovering}. In early 2021, there was a significant surge in the research and development related to vaccine production to build antibodies to fight against the COVID-19. AstraZeneca, Pfizer, Sputnik, ZyCoV-D, Covieshield, and Covaxin are, to name a few, with high efficacy. However, as the COVID-19 keeps mutating into various other forms, it becomes more challenging for the vaccines to neutralize them.

\begin{figure*}[t!]
 \centering
  \includegraphics[scale=0.90]{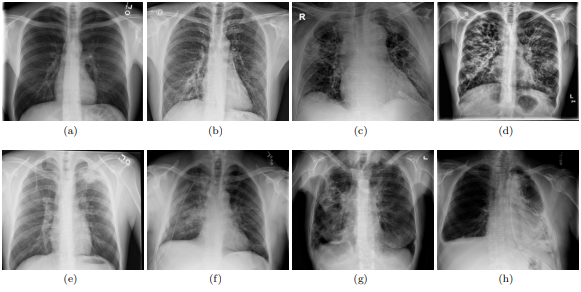}
  \caption{Sample Chest X-Ray images from the COVIDx \cite{lindawangg} dataset (a) Normal (b) COVID-19 - Mild infection (Chest X-ray shows non-homogenous ground glass opacities in the right paracardiac region in the mid and lower zones.) (c) COVID-19 - Moderate infection (Bilateral ground glass and alveolar opacities ) (d) COVID-19 - Severe infection (Severe Bilateral extensive involvement with ground glass, alveolar, interstitial shadows with cavitation involving right lung more than the left lung)   (e) Pneumonia - mild infection (Chest X-ray shows an dense alveolar opacity in the left upper lobe with small cavitation suggesting necrotising pneumonia.) (f) Pneumonia - moderate infection (Moderate, alveolar opacities in the right mid zone extending to upper and lower zones with 2 ICD’s in situ) (g) Pneumonia - Severe infection (Severe, Bilateral opacities extending to five lung zones, silhouette of both cardiac borders, possible mastectomy, lung fibrosis) (h) Pneumonia - Severe infection (Severe, Necrotising pneumonia with cavitation and lung abscess in left lower zone, extending to left mid and upper zone. Loss of volume evident in the left lung. Mild Pleural effusion right.)}
\label{fig:CXRR}
\end{figure*}

The definitive method of diagnosing COVID-19 involves the usage of a swab from the nose and throat from the patient and performing a real-time Reverse Transcription-Polymerase Chain Reaction (RT-PCR) test to detect the viral RNA strand \cite{tang2020laboratory}. In addition, FDA has permitted many laboratory-developed tests for emergency usage. Though these methods demonstrate high analytic sensitivity and specificity in ideal settings, the clinical performance depends on the type and quality of the specimen and the duration of illness at the testing time. However, these methods are subjected to high false negativity rates ranging from 5-40\%. Nevertheless, the turnaround time varies from 15 minutes to 8 hours, depending on the test and laboratory workflow in a busy setting \cite{cheng2020diagnostic,weissleder2020covid}. Having understood the shortcomings of RT-PCR or laboratory testing and the highly contagious nature of COVID-19, there is a need for a test that can give a faster diagnosis and enable early isolation. Though less sensitive than CT, in the account of public health concern, chest X-Ray (CXR), when abnormal, can be an essential tool in suspecting a case of COVID-19.  It also helps in triaging, risk stratification, early management, and monitoring the progress in moderate to severe cases while RT-PCR reports are awaited.  Though the findings of CXR in COVID-19 are not specific, a few of the classical findings such as bilateral, basal, and peripheral ground-glass attenuation with occasional consolidation helps to differentiate it from non-COVID-19 pneumonia, and healthy cases. Fig.~\ref{fig:CXRR} depicts the sample Chest X-Ray images with varying severity levels. Despite its limited sensitivity in detecting early disease, nearly 70\% of those requiring hospitalization have abnormal radiographs that progress to peak in about 10-12 days after symptom onset. Chest radiographic abnormalities are typically air space opacities (either ground-glass opacities or consolidation), which are most often bilateral and peripheral with lower lobe predominance. Thus, the analysis of the CXR as an alternative tool for diagnosing and managing COVID-19 has gained interest. 

\begin{figure*}[t!]
 \centering
  \includegraphics[scale=0.11]{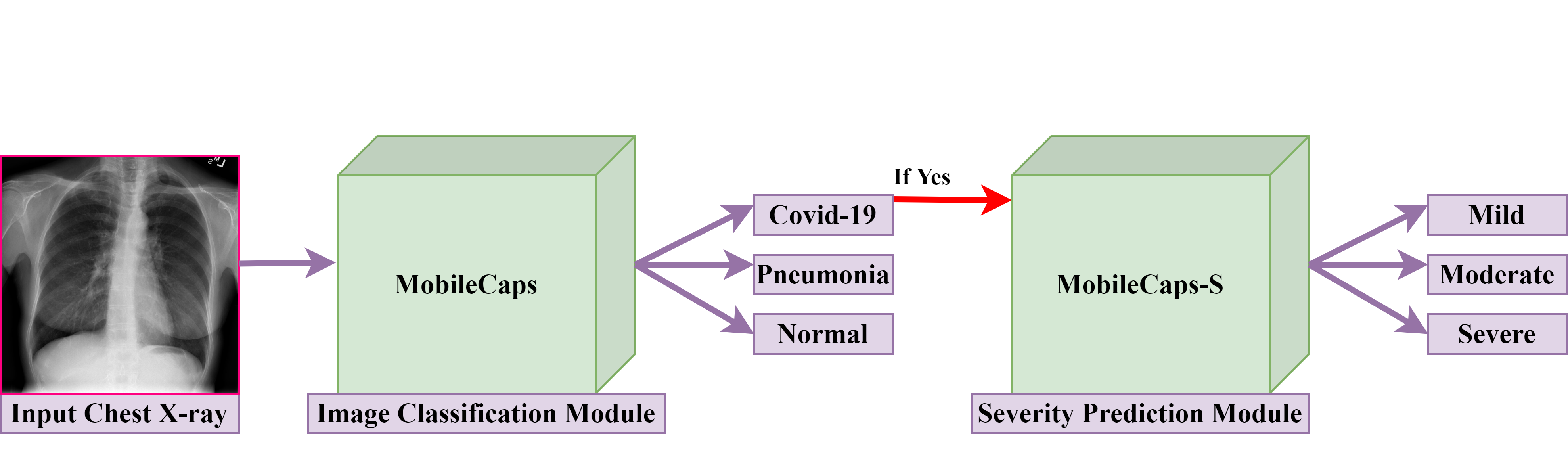}
  \caption{The workflow of the proposed framework for the severity assessment of COVID-19 CXR.}
  \label{fig:pipe}
\end{figure*}

The accuracy of CXR diagnosis heavily relies on radiological expertise because of the complex patterns involved, which is a huge disadvantage because of the limited number of expert radiologists, particularly in developing countries. Automating the process can reduce the workload of the radiologists/pulmonologists and can provide fast and appropriate treatment to the patients. Apart from detecting the abnormality from the CXR, assessing the severity of infection based on the severity scoring metrics such as BRIXIA and RALE has also gained attention in the research community. Fig.~\ref{fig:pipe} depicts the workflow of the proposed severity assessment model using COVID-19 CXR. Several automated methods were proposed in the literature for detecting CXR with COVID-19 abnormality due to the availability of datasets  \cite{cohen2020, Chung, Chungg, North, NNorth, Tweet}. However, there are minimal works been appeared in the literature discussing the severity assessment. The main contributions in this paper are summarized below:
\begin{itemize}
    \item[$\bullet$] We introduce MobileCaps, a novel hybrid architecture that integrates MobileNetV2 and Capsule Networks for efficient  classification of COVID-19 CXR images. The proposed architecture significantly reduces the number of trainable parameters while giving better performance as compared to state-of-the-art models.
    \item[$\bullet$] We demonstrate that the proposed model is highly generalizable as compared to other studies in the literature by extensively evaluating on the diverse COVIDx test set, which is an amalgamation of five publicly available datasets. We also evaluated the proposed model on two other RT-PCR confirmed data sets to demonstrate the generalizability and superiority of the proposed model over the state-of-the-art.
    \item[$\bullet$] We propose MobileCaps-S for analyzing the severity of the COVID-19 infection. The proposed model can predict the severity based on the Radiographic Assessment of Lung Edema (RALE) scoring technique. 
\end{itemize}
The rest of the paper is organized as follows, Section 2 discusses the automated methods proposed in the literature for classifying CXR images of COVID-19 from non-COVID-19 Pneumonia and normal cases. The proposed method is outlined in Section 3. Experimental results and the discussions are presented in Section 4 and finally, we conclude the paper in Section 5.

\section{Literature Survey}
Automated screening of COVID-19 using Chest X-Ray is gaining a lot of significance in the battle against the pandemic. Several studies have been proposed in the literature to develop a fully automated AI-based solution for screening COVID-19 CXR images. Automated screening techniques can be categorized into two categories based on the tasks: 1) Classification 2) Severity score prediction. Classification tasks can be further classified into Two-class classification (COVID-19 and non-COVID-19), Three-class classification (COVID-19, non-COVID-19 Pneumonia, and Normal cases), and Four-class classification (COVID-19, Bacterial Pneumonia, non-COVID-19 Pneumonia, and Normal cases).

Two-class classification involves classifying COVID-19 from non-COVID-19 cases, including CXR with other infections such as SARS, MERS, ARDS, along with normal cases. Afshar et al. \cite{afshar2020covid} introduced COVID-CAPS, a Capsule Network-based framework for the classification of COVID-19 CXR.  Sethy et al.\cite{sethy2020} presented a CNN model coupled with a Support Vector Machine (SVM) classifier for the defined objective. The intent behind both of the above works is to differentiate COVID-19 cases from non-COVID-19. Abraham and Nair \cite{abraham2020computer} introduced a multi CNN approach, which involves s series of pre-trained CNN models for accurately screening COVID-19 CXR. This approach aggregates the features extracted by the multiple CNN models with the Bayesnet classifier and correlation-based feature selection (CFS) technique to achieve the objective. The authors have evaluated the performance of the proposed method on two publicly available datasets and achieved a reasonably good performance. Horry et al. \cite{horry2020covid} used a transfer learning approach with VGG 19 architecture (initialized with IMAGENET weights) to effectively detect COVID-19 radiographs. In the study, the authors have experimented with three modalities, including CT, Ultrasound, and CXR, and have claimed ultrasound images delivered a good detection rate. However, to develop a reliable system, the model should be capable of handling COVID-19 cases in a multi-class scenario, which turns to be the major drawback of this approach.

The majority of the works proposed in the literature focused on three-class classification involving the identification of COVID-19 cases from non-COVID-19 Pneumonia and normal cases. A semi-supervised hybrid model was proposed by Khobahi et al. \cite{khobahi2020coronet} by using Task-Based-Feature-Extraction-Technique (TFEN) followed by COVID Identification Network (CIN). However, the study was conducted with minimal samples of COVID-19 cases (89 and 10 samples in training and testing, respectively), which turned out to be the major pitfall of this approach. COVIDiagnosis-Net was introduced by Ucar et al. \cite{ucar2020covidiagnosis} by using SqueezeNet architecture with Bayesian Optimization and an offline augmentation technique. In this work as well, the model was evaluated on limited samples of COVID-19 cases (66 and 10 samples for training and testing, respectively). Though the authors extensively used data augmentation techniques to increase the COVID-19 samples (1229 and 154 samples for training and testing, respectively), it raises concerns as to what extent these techniques will translate to real-world data variability making generalizability the major hindrance of this approach. Eduardo et al. \cite{luz2020towards} demonstrated the use of EfficientNet with Mobile Inverted BottleNet Conv (MBConv) and data augmentation for classifying COVID-19 CXR images. This model exhibits two potential limitations, 1) Poor localization on class activation maps 2) A limited number of COVID-19 samples (152 and 31  samples for training and testing, respectively). Rahimzadeh et al. \cite{rahimzadeh2020modified} introduced a modified CNN that concatenates the features extracted from Xception and ResNet50V2 architectures to form a rich feature space for the accurate detection of CXR images of COVID-19 cases. This model achieved a recall rate of 80.53 and a precision of 35.27 for COVID-19 cases.

Xin et al. \cite{li2020covid} introduced an architecture named COVID MobileXpert for screening COVID-19 CXR using the Knowledge Transfer Distillation framework (KTD) and Attending Physician network (AP). However, the results were reported on a severely less number of COVID-19 CXR samples (125 samples for training and 36 samples for testing). DarkCovidNet architecture presented by Tulin et al. \cite{ozturk2020automated} used You Look Only Once (YOLO) architecture for the quick assessment of COVID-19 CXR. Poor localization and evaluation with minimal samples of COVID-19 cases (129 samples) turn to be a major downfall of this approach. Karim et al. \cite{karim2020deepcovidexplainer} proposed DeepCOVIDExplainer, a visually guided residual network-based CNN architecture for the classification of COVID-19 CXR images. Authors applied pre-processing techniques such as edge enhancement and denoising to improve the quality of input images. DeepCOVIDExplainer was trained with the snapshot neural ensemble technique and achieved a recall rate and precision of 88.1 and 87.7 respectively on unbalanced data and 90.5 and 90.4 on balanced data for COVID-19 cases. Nour et al. \cite{nour2020novel} introduced a combination of CNN-based models and a machine learning classifier (SVM) to diagnose COVID-19 radiographs from normal and pneumonia cases. The authors have used a Bayesian optimizer for tuning the parameters of machine learning models. This method achieved an accuracy of 98.97, F1 score of 96.72 with a lower sensitivity of 89.39.

Wang et al. \cite{wang2020covid} formulated a public dataset called COVIDx \cite{lindawangg}, which is a combination of many publicly available datasets \cite{cohen2020,Chung,Chungg,North,NNorth}  and also proposed an architecture called COVID-Net which makes extensive use of Projection-Expansion-Projection (PEP) blocks for efficient classification of COVID-19 CXR images from normal and non-COVID-19 Pneumonia cases. The model achieved an overall accuracy of 93.3 and a precision and recall rate of 98.9 and 91.0, respectively, for COVID-19 cases. Oh et al. \cite{oh2020deep} introduced CNN with a patchwise wise approach based on the statistical interpretation of CXR data. Authors showed interesting results in comparison with \cite{wang2020covid} by achieving an overall accuracy of 91.9 with the recall rate of 100 and low precision of 76.9 for COVID-19 cases. A four-class classification approach is followed by Farooq et al. \cite{farooq2020covid} involving the detection of COVID-19 cases from Normal, non-COVID-19 Pneumonia, and Bacterial Pneumonia cases using residual blocks. This study was carried out with limited samples of COVID-19 cases (68 samples), due to which it lacks generalizability. Aslan et al. \cite{aslan2021cnn} proposed a transfer learning-based hybrid architecture for accurately classifying COVID-19 radiographs from normal and pneumonia cases. This method combines AlexNet architecture and BiLSTM for incorporating temporal information to achieve the defined objective. This model recorded an overall accuracy of 98.70.

Severity assessment or severity score prediction involves assessing the progression of COVID-19 infection by observing and the lung involvement through the standard scoring techniques such as BRIXIA \cite{borghesi2020covid}, or Radiographic Assessment of Lung Edema (RALE) \cite{warren2018severity} scoring systems. BRIXIA score was an experimental scoring system developed by researchers from Italy during the early COVID-19 pandemic. In the BRIXIA scoring system,  both lungs are divided separately into three different zones and classified each zone into 4 different severities based on lung opacities.  These opacities though not specific, are characteristic of viral Pneumonia of COVID-19. In literature, we can observe few methods based on the BRIXIA scoring system to predict the severity of COVID-19 infection \cite{signoroni2020end, borghesi2020chest,setiawati2021modified}. RALE scoring system was initially developed as a non-invasive measure to assess the severity and outcome profile of the patient with ARDS \cite{warren2018severity}. As ARDS is one of the features of Severe COVID pneumonia, modified RALE score, especially the geographic assessment, was found useful to assess the severity of COVID-19 Pneumonia in many studies. The scale of RALE score ranges from 1-8 (1-2: Mild, 3-5: Moderate, 6-8: Severe). Radiologists can assess this score to measure the severity and to treat the patients accordingly. Authors of \cite{wang2020covid} introduced COVIDNet-S, a deep neural network model for predicting the geographic extent and opacity extent of COVID-19 CXR. COVIDNet-S achieved an R$^2$ score of 66.4 and 63.5 for geographic extent and opacity extent, respectively. Tabik et al. \cite{tabik2020covidgr} introduced COVID-Smart Data-based Network (COVID-SDNet) for classifying the infection progression into mild, moderate, and severe based on the RALE scoring technique. Apart from the BRIXIA and RALE scoring, there are some other techniques that are being adopted in the literature for severity scoring of COVID-19 CXR infection \cite{baratella2020severity,cohen2020predicting, reeves2020performance,irmak2021covid}. 

\section{Methods}
Convolutional Neural Networks (CNN) have revolutionized the computer vision domain with their ability to learn the intrinsic features of an image by using a gradient-descent-based learning algorithm, at times even surpassing human-level performance on many public datasets. Despite their massive success, certain conditions need to be met for a CNN to perform well. Firstly, a CNN requires the availability of huge amounts of data, which is hard to come by in medical imaging. Secondly, a typical CNN incurs a very high computational cost due to its large number of trainable parameters. Moreover, a CNN does not take into account spatial hierarchies among the features \cite{sabour2017dynamic}, which are paramount in high-level image encoding/understanding. On the other hand, Capsule networks follow a different approach and address the above flaws with the help of a powerful neuronal representation called \textit{capsules} and the dynamic routing algorithm. The first stage of Capsule Networks generally follows a series of simple convolutional layers to form primary capsules for initial level feature encoding. We demonstrate and show that using an efficient and powerful organization of convolutional layers instead of a series of simple convolutional layers would lead to powerful primary capsules. To this end, we propose a lightweight hybrid model called MobileCaps to build a rich feature space while requiring a fraction of the computational cost compared to other CNN models. We use MobileNetV2 for the initial stage of MobileCaps due to its high efficacy with a smaller parameter space, followed by capsule layers.  In the following sub-section, we have brieﬂy elaborated on the Capsule Networks and MobileNetV2, which follows the proposed model.

\subsection{Capsule Networks}
Sabour et al. \cite{sabour2017dynamic} introduced Capsule Networks with an intent to address the limitations of CNNs, namely the large data requirement and their inability to learn hierarchical spatial relationships. Capsule Networks comprise multiple capsules that hold a vector value instead of a neuron that holds a scalar.   Each capsule consists of instantiation parameters of the features of an image, and the length of a capsule represents the probability that the feature exists in the image. Capsules communicate with each other using the dynamic routing algorithm. A capsule $u_{i}$ in layer $l$ will try to predict the output of another capsule $\hat{u}_{j|i}$ in layer $l+1$ using a trainable weight matrix $W_{ij}$ as given in Eq \ref{eq1} \cite{sabour2017dynamic}:
\begin{equation}
    \hat{u}_{j|i} = {W}_{ij}\cdot{u}_{i} \label{eq1}
\end{equation}
Each capsule $u_{i}$ has a \textit{routing coefficient} $c_{ij}$ associated with it that corresponds to how much it \textit{agrees} with the output of the capsule $u_{j}$. The value $c_{ij}$ is determined by the dynamic routing algorithm. The actual output $s_j$ of capsule $u_j$ is calculated as shown in Eq \ref{eq2} \cite{sabour2017dynamic}. 
\begin{equation}
    {s}_{j} = \sum_{i} {c_{ij}}\cdot{\hat{u}_{j|i}} \label{eq2}
\end{equation}
The output vector $s_j$ is then passed through a non-linear squash activation as can be seen in Eq \ref{eq3}. This is done to preserve it's direction by restricting it's length to the range $[0,1]$ to obtain the final output $v_j$.
\begin{equation}
    {v}_{j} = \frac{{||s_{j}||}^2}{1+{||s_{j}||}^2}\cdot \frac{s_{j}}{{||s_{j}||}} \label{eq3}
\end{equation}
The similarity between the actual output $v_j$ and the predicted output $\hat{u}_{j|i}$ is calculated by taking their dot product, i.e. a measure of how much the predicted output \textit{agrees} with the actual output is computed and the routing coefficient for each capsule $u_i$ is updated as per Eq \ref{eq4} and Eq \ref{eq5} \cite{sabour2017dynamic}. This is the essence of the dynamic routing algorithm.
\begin{equation}
    {b}_{ij} = {b}_{ij} + {v}_{j}\cdot\hat{u}_{j|i} \label{eq4}
\end{equation}
\begin{equation}
    {c}_{ij} = \frac{exp(b_{ij})}{\sum_k exp({b_{ik}})} \label{eq5}
\end{equation}
Initially all capsules $u_i$ are given equal weightage. Eqs \ref{eq2}-\ref{eq5}  can be repeated $r$ times to obtain a better value of their routing coefficients, where $r$ indicates the number of \textit{routing iterations}.
\begin{figure*}[t!]
 \centering
  \includegraphics[scale=0.060]{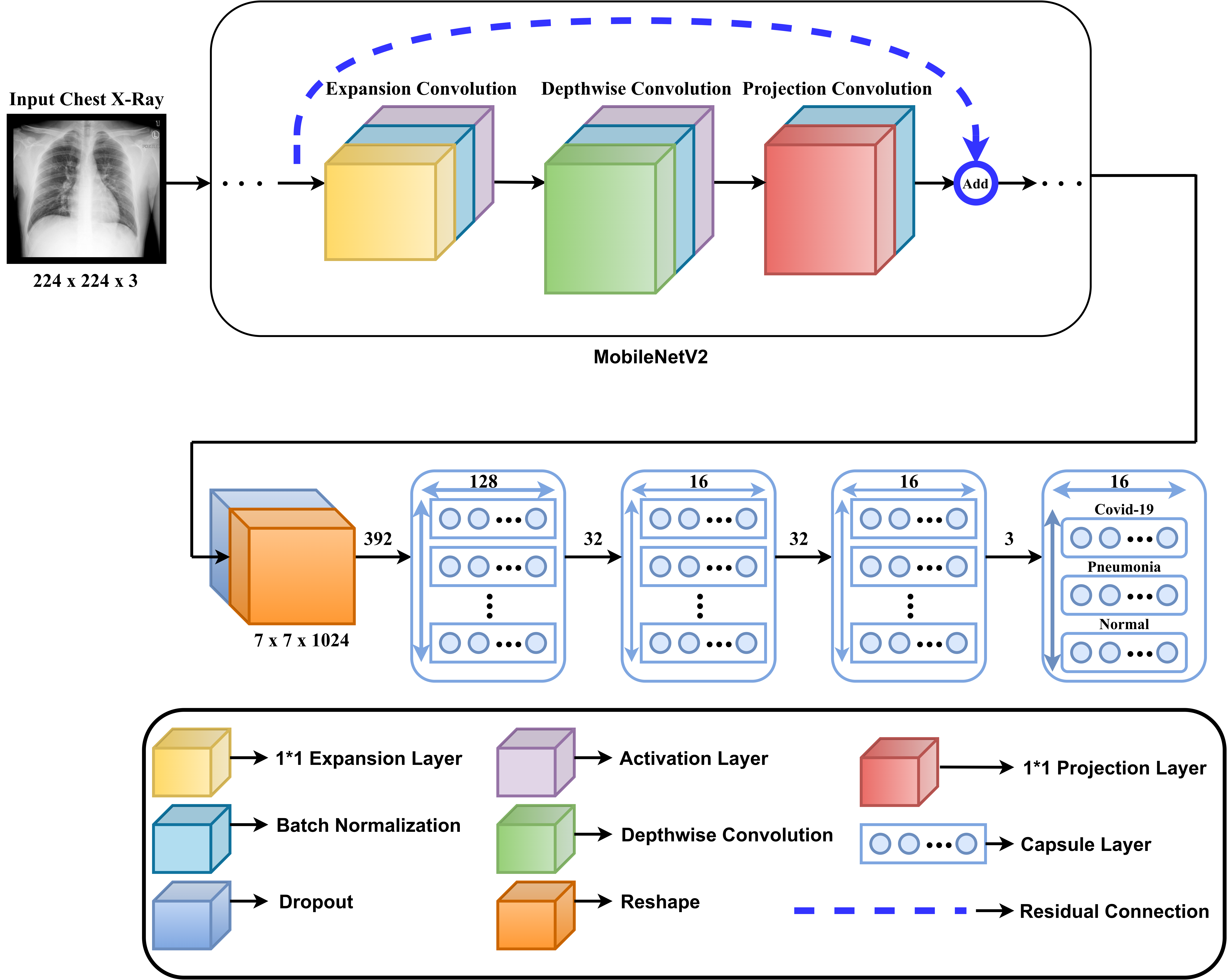}
  \caption{Proposed MobileCaps Architecture.}
  \label{fig:arch}
\end{figure*}
\subsection{MobileNetV2}
MobileNetV2 \cite{sandler2018MobileNetv2} is a family of lightweight CNN models that are suitable for mobile and embedded computer vision applications.  MobileNetV2 aims at replacing expensive convolution operations with less expensive and efficient ones without compromising on performance. The factorization of a standard convolution into a depth-wise convolution followed by $1\times 1$ convolution (depth-wise separable) with inverted residuals and linear bottlenecks forms the crux of the architecture. It majorly consists of three blocks, namely the expansion block, the depth-wise convolution block, and the projection block, respectively.

The expansion block is a $1\times1$ convolution block that accepts the input tensor and expands the number of channels specified by the \textit{expansion-factor}. Expansion-factor is a hyperparameter to be set (default value is 6). For example, If the expansion block accepts an input tensor with 32 channels, it will output a new tensor with 192 (32$\times$6) channels. The resultant output is then filtered using a depth-wise convolution block and fed into the projection or the bottleneck layer, which is also a $1\times1$ convolution layer that shrinks or projects the input back to a smaller number of channels (say, 192 to 32). MobileNetV2 also makes use of residual connections to boost the gradient flow throughout the network. 

\subsection{Proposed Model}
The proposed model is shown in Fig.~\ref{fig:arch}. It is an amalgamation of MobileNetV2 and Capsule Network that results in a lightweight and efficient, hybrid architecture wherein MobileNetV2 extracts features from the CXR image, which is subsequently passed to the Capsule Network. Generally, the first stage of Capsule Networks is a series of convolutional layers meant for extracting and encoding simple features like edges, horizontal lines, vertical lines, etc., in the form of a vector. The output of the final convolution operation is reshaped to form primary capsules that communicate with higher-level capsules with the help of the dynamic routing algorithm (as described from Eqs \ref{eq1}- Eq \ref{eq5}). We observed that making the initial set of convolutional layers deeper and more efficient would help build better primary capsules, which gives an excellent start to the capsule network, thereby improving the model's overall performance. Our observations are corroborated by the ablation study, which we discuss in the following subsection.

MobileCaps accepts Chest X-ray (CXR) images of shape $224 \times 224 \times 3$, which is fed into the first block i.e a MobileNetV2 that has been initialized with ImageNet weights \cite{deng2009imagenet}. The MobileNetV2 acts as a feature extractor that constructs a rich feature space of dimension $7 \times 7 \times 1024$. This feature map is then passed through a dropout layer to enforce regularization and to prevent possible overfitting. This output is then reshaped into the primary capsule layer of dimension $392 \times 128$, i.e., 392 capsules of dimension 128. This layer is followed by two more capsule layers of dimension $32 \times 16$ that use the dynamic routing algorithm (Eqs. \ref{eq2}-\ref{eq5}) to pass information from one layer to the next. The final capsule layer contains three 16-dimensional capsules, one for each class. The prediction vector is calculated by computing the length of these capsules, which indicates the probability of the presence of each class. We further propose MobileCaps-S for predicting the severity score of COVID-19 CXR based on the RALE score. We modified the bottom layers of the proposed MobileCaps architecture for the defined objective. The model accepts the Chest X-Ray image of the shape $224 \times 224 \times 3$  and predicts the probability output in the range of 0 to 1, indicating the severity, which is then mapped back into the RALE severity score (1-8). The number of capsule layers and the number of capsules in each capsule layer are fine-tuned to get the optimal performance with minimal parameter space. In the next subsection, we provide observations of the ablation study conducted and drive home the intuition behind using  MobileNetV2 as a feature extractor for the Capsule Network.

\begin{table*}
\centering
\renewcommand{\arraystretch}{1.4}
\caption{Ablation Study.}
\begin{tabular}{@{}lcccccccccccccc@{}}\toprule
Models & 

\multicolumn{3}{c}{\textbf{Precision}} & \multicolumn{3}{c}{\textbf{Recall}} & 
\multicolumn{3}{c}{\textbf{F1 Score}}\\

\cmidrule(lr){2-4} \cmidrule(lr){5-7} \cmidrule(lr){8-10} 
& \textbf{C} & \textbf{N} & \textbf{P}   & \textbf{C} & \textbf{N} & \textbf{P}  &\textbf{C} & \textbf{N} & \textbf{P}\\ \midrule
CapsNet  & 67.25 & 85.24 & 78.43  & 75.27 & 82.64  & 83.20 & 71.03 & 83.91 & 80.54 \\
MobileNetV2  & 84.16 & 87.81 & 79.74 & 83.60 & 77.00 & 85.20 & 83.87& 82.05 & 82.37 \\
\textbf{Proposed Method} & \textbf{98.50} & \textbf{88.21} & \textbf{92.62}  & \textbf{91.60} & \textbf{94.60} & \textbf{92.20} & \textbf{94.91} & \textbf{91.22} & \textbf{92.33}\\

\bottomrule
\end{tabular}
\label{Table 2}
\end{table*}

\subsection{Ablation Study}
As a part of this study, we performed an ablation analysis supporting the design of the proposed architecture. We compared the performance of the proposed model by training Capsule Networks and MobileNetV2 independently. As per our hypothesis, the performance of Capsule Network that uses simple convolutions to form primary capsules is subpar. Moreover, since MobileNetV2 misses out on the advantage of capsules and the dynamic routing algorithm, it performs poorly compared to the proposed architecture. Table \ref{Table 2} shows the quantitative results depicting the superiority of the proposed architecture. Capsule Network and MobileNetV2 achieved an F1 score of 74.03 and 82.90, respectively whereas, the proposed method achieved 94.91, which is remarkable. The following section presents the experimental results and discussion of our proposed model.
\begin{table}[hbt!]
\centering
\caption{Train/Test Distribution of Different Datasets Adopted in the Study.}
\begin{tabular}{p{4.0cm}cccc}\toprule
\textbf{Dataset} & \textbf{Type} & \textbf{COVID-19} & \textbf{Normal} & \textbf{Pneumonia} \\
\midrule
COVIDx \cite{lindawangg} & Train & 473 & 7966 & 5459 \\
& Test & 100 & 594 & 885 \\
&&&& \\
COVIDx Official\cite{lindawangg} & Train & 473 & 7966 & 5459 \\
& Test & 100 & 100 & 100 \\
&&&& \\
COVIDx Balanced\cite{lindawangg} & Train & 1500 & 1500 & 1500 \\
& Test & 100 & 100 & 100 \\ 
&&&& \\
Twitter COVID-19 Data\cite{Tweet}  & Test & 134 & - & - \\
&&&& \\
JSS COVID-19 Data & Test & 76 & - & - \\
\bottomrule
\label{Table 3}
\end{tabular}
\end{table}
\section{Experimental results and Discussion}

This section epitomizes the datasets used in the study, followed by the training strategy, loss function, and hyperparameters adopted in the proposed method. Finally, we evaluate and compare the quantitative performance of the proposed model with various other benchmark models.

\subsection{Dataset Description}
In this subsection, we will briefly elaborate on the different datasets used in the study. We consider the recent version of the COVIDx dataset (accessed 13 June 2020) \cite{lindawangg} introduced by Wang et al.\cite{wang2020covid}, Twitter COVID-19 CXR data \cite{Tweet}, JSS COVID-19 data, and COVIDx RALE severity data for evaluating the performance of the proposed model.
\begin{itemize}
\item[$\bullet$] \textbf{COVIDx data \cite{lindawangg}:} COVIDx data  is formed by merging the following publicly available datasets. 1) COVID-19 image data collection \cite{cohen2020} 2) COVID-19 chest x-ray data initiative \cite{Chung} 3) Actualmed COVID-19 chest x-ray data initiative \cite{Chungg} 4) COVID-19 radiography database \cite{North} 5) RSNA Pneumonia Detection Challenge dataset  \cite{NNorth}. COVIDx \cite{lindawangg} is widely considered as the benchmark dataset for developing an automated algorithm for accurate classification of COVID-19 CXR from non-COVID-19 CXR and healthy cases. The data distribution of actual and the official version of the COVIDx data adopted by the authors of COVIDx \cite{lindawangg} is shown in the first and second rows of Table \ref{Table 3}. Further, COVIDx \cite{lindawangg} data suffers from data imbalance, which can be observed in Table \ref{Table 3}. The number of samples in the COVID-19 class is less as compared to the Normal and non-COVID-19 Pneumonia classes. To mitigate the challenge of data imbalance,  we over-sampled the COVID-19 class by applying data augmentation techniques (horizontal flip, zoom, rotation, width shift, height shift, and shear). We increased the size of the COVID-19 class to 1500 samples while randomly under-sampling the other two classes to obtain 1500 samples from each class, and thus we prepared and adopted a balanced version of COVIDx data in this study. 
\item[$\bullet$] \textbf{Twitter COVID-19 CXR data \cite{Tweet}:} This dataset consists of 134 Chest X-Rays manifesting COVID-19 viral pneumonia, which is made available on Twitter for the research purpose by a radiologist from Spain. All 134 cases are subjected and confirmed with SARS-CoV-2 PCR+.
\item[$\bullet$] \textbf{JSS COVID-19 CXR data:} This data is obtained from Jagadguru Sri Shivarathreeshwara (JSS) medical college Mysore, India. This dataset consists of 76 COVID-19 RT-PCR confirmed cases from 76 patients.
\item[$\bullet$] \textbf{COVIDx RALE severity data:}  COVID-19 CXR from COVIDx data \cite{lindawangg} was scored with the RALE severity scoring method \cite{warren2018severity} with the help of a senior radiologist and a pulmonologist using the Labelbox \cite{label} annotation tool.  There were a total of 573 samples,  but due to data imbalance (a large number of mild and moderate cases as compared to the number of severe cases), we have considered a total of 388 CXR with a varying severity level.
\end{itemize}

\begin{figure*}[t!]
 \centering
  \includegraphics[scale=0.5]{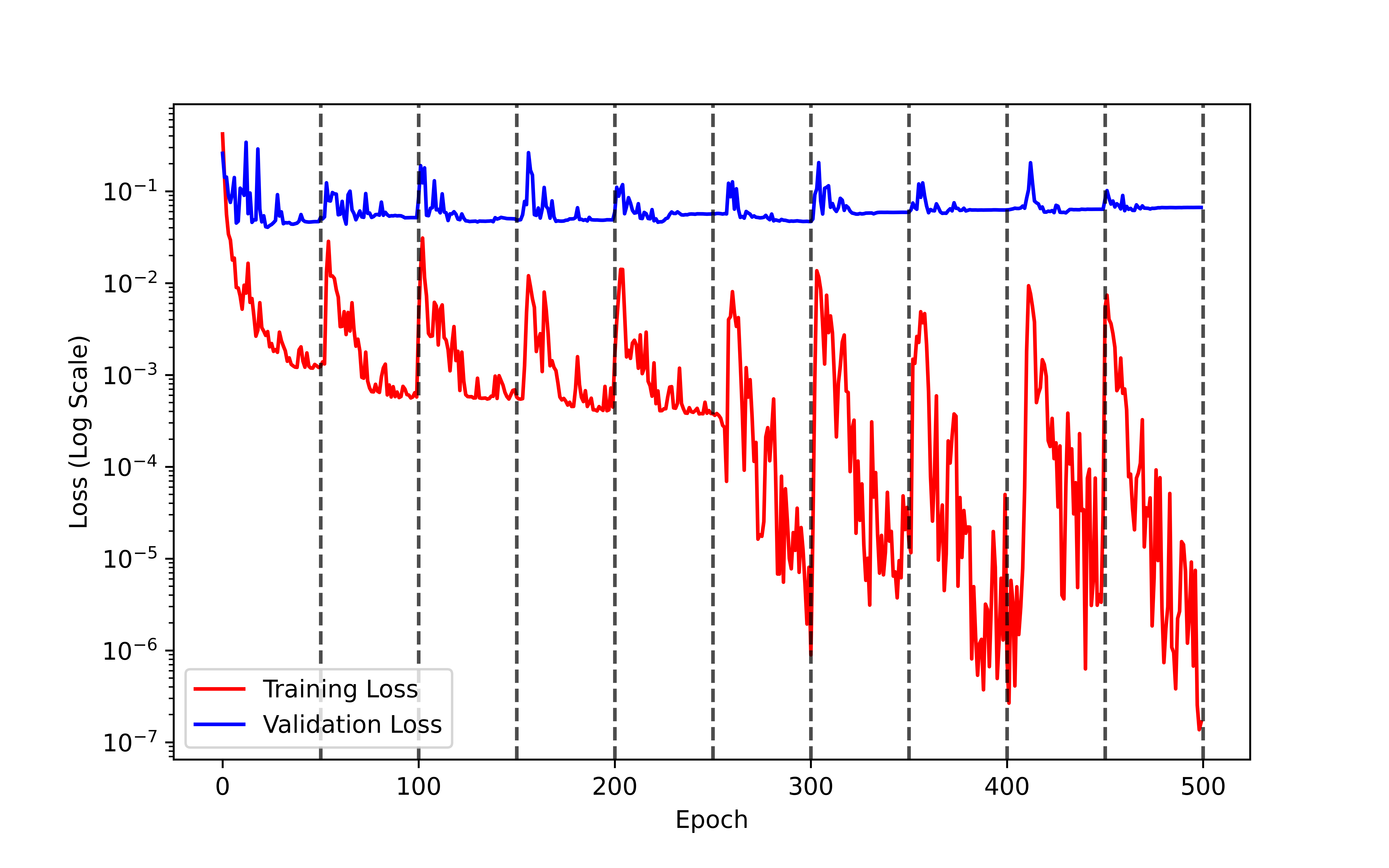}
  \caption{The Learning Curve of Proposed Method.}
  \label{fig:loss1}
\end{figure*}

\subsection{Training Methodology}
We train the proposed model on the preprocessed COVIDx \cite{lindawangg} dataset with a cyclic cosine annealing policy \cite{loshchilov2016sgdr} for the learning rate along with a snapshot ensembling strategy \cite{huang2017snapshot}. The policy starts with a large learning rate that aggressively decreased to a minimum value before increasing again. The learning rate is changed according to Eq \ref{eq9} \cite{loshchilov2016sgdr}. Fig.~\ref{fig:loss1} shows the learning curve of the proposed method.

\begin{equation}
    \alpha(t) = \frac{\alpha_0}{2}\left( \cos \left( \frac{\pi \, \text{mod} \left( t-1, {\lceil T/C \rceil} \right)}{\lceil T/C \rceil}\right)+1 \right)
    \label{eq9}
\end{equation}
where $\alpha_0$ denotes the maximum learning rate, $t$ is the current epoch number, $T$ is the total number of epochs, and $C$ is the total number of cycles. The decaying learning rate during the cycle allows the model to converge to a local minimum. At the end of the cycle, the high value of the restarted learning rate is enough to perturb the model from its local minima, following which the decaying learning rate allows the model to converge to another local minimum. We take snapshots of the model at each such local minima, as there is significant diversity in the local minima visited by the model.
With the snapshot ensembling method, we can get $M$ models while incurring the computational cost of training only one model. During testing, we combine the results of $M$ snapshots by taking the average of the softmax probabilities of each snapshot. The initial learning rate and the length of one cycle are hyperparameters, the optimal value for which is determined by the Bayesian optimization technique\cite{loshchilov2016sgdr}. 

Bayesian optimization makes informed decisions about the next set of hyperparameters to evaluate based on the previous results. It takes fewer iterations for the algorithm to find the optimal combination of hyperparameters as it disregards those values that won't improve the validation score. It achieves this by finding an approximation for the objective function called the surrogate function that can be sampled effectively.  We use this technique to find an optimal value for the learning rate and the cycle length. For each Bayesian optimization process's estimate, we train our proposed model for one cycle to obtain a reasonable estimate of the validation accuracy. We select the model with the best validation accuracy and train it for 10 cycles, taking a snapshot at the end of each cycle, which is then used to make the final prediction.

We train the proposed model till convergence, which approximately took 500 epochs after selecting the best hyperparamaters with Bayesian optimization. Nadam optimization \cite{dozat2016incorporating} with $\beta_1 = 0.9$, $\beta_2 = 0.999$, and $\epsilon = 10^{-7}$
was used along with snapshot ensembling with snapshots being taken every 50 epochs. MobileCaps was trained with the margin loss function described in Eq \ref{eq12} \cite{sabour2017dynamic} and MobileCaps-S was trained with logcosh loss function as describe by Eq \ref{eq13} respectively.

\begin{equation}
\begin{aligned}
    L_k =& T_k\,\text{max}\left(0,m^+ - ||\textbf{v}_k||\right)^2 + \\ &\lambda\,\left(1-T_k\right)\,\text{max}\left(0,||\textbf{v}_k|| - m^- \right)^2
    \label{eq12}
\end{aligned}
\end{equation}
where $\textbf{v}_k$ denotes the final set of capsules, and $T_k=1$ only if the $k'th$ class is present in the image. $m^+$, $m^-$, and $\lambda$ are hyperparameters whose values are 0.9, 0.1 and 0.5 respectively.

\begin{equation}
\begin{aligned}
           L(y,y^p) = \sum_{i}^ n log(cosh(y^p_i-y_i))
          \label{eq13}
\end{aligned}
\end{equation}

\subsection{Results and Discussions}
In this section, we will quantitatively analyze the performance of the proposed MobileCaps method on the balanced COVIDx data \cite{lindawangg}, Twitter COVID-19 CXR data \cite{Tweet} and the JSS COVID-19 CXR data and MobileCaps-S on COVIDx RALE severity data. The classification task is evaluated with precision, recall, and F1-score, and the severity prediction uses R$^2$ coefficient as the evaluation metric.

\begin{figure}
    \centering
    \subfloat[\centering Confusion Matrix]{{\includegraphics[width=8.5cm]{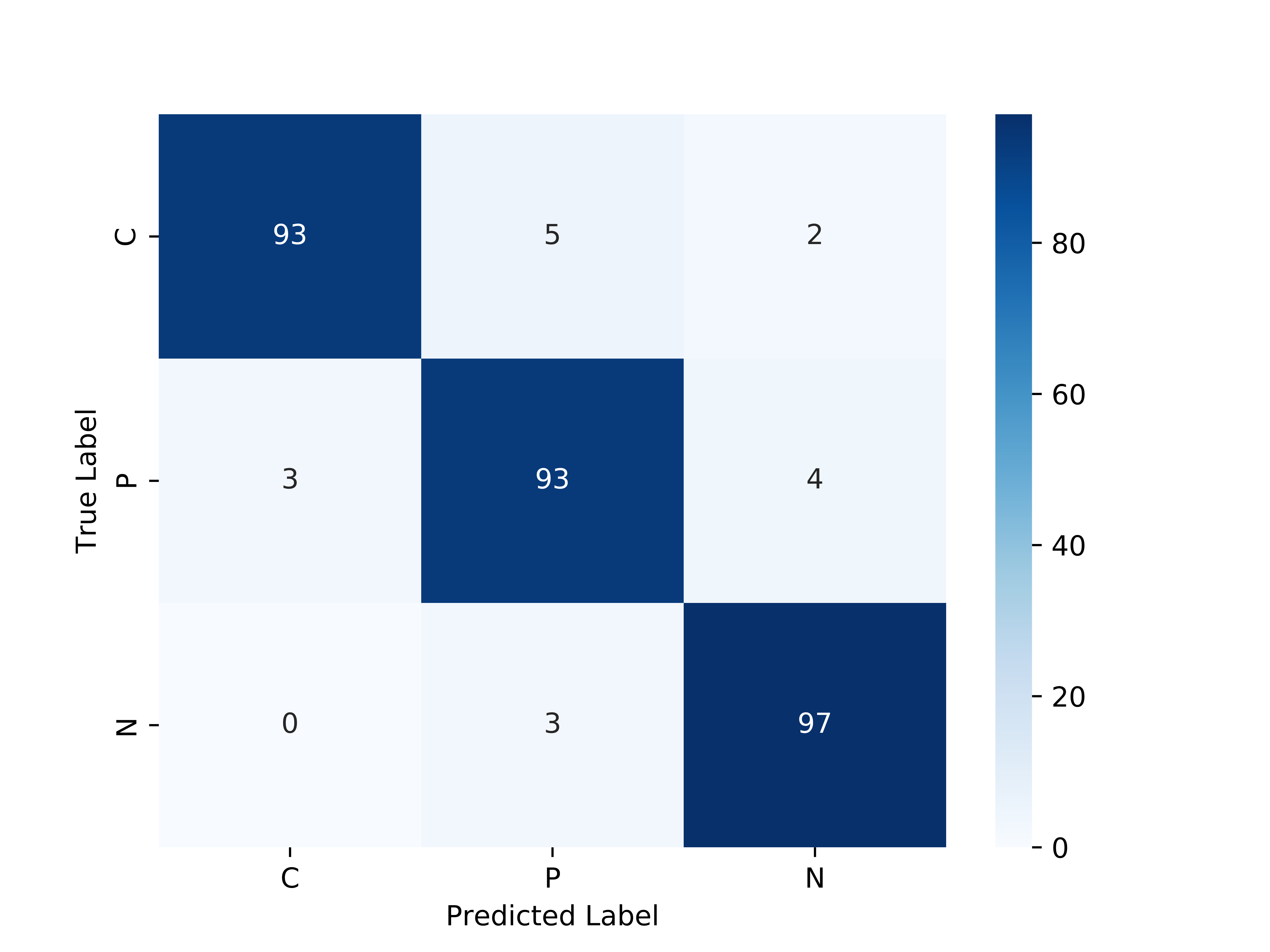} }}
    %\qquad
    \subfloat[\centering Receiver operating characteristic curve.]{{\includegraphics[width=8.5cm]{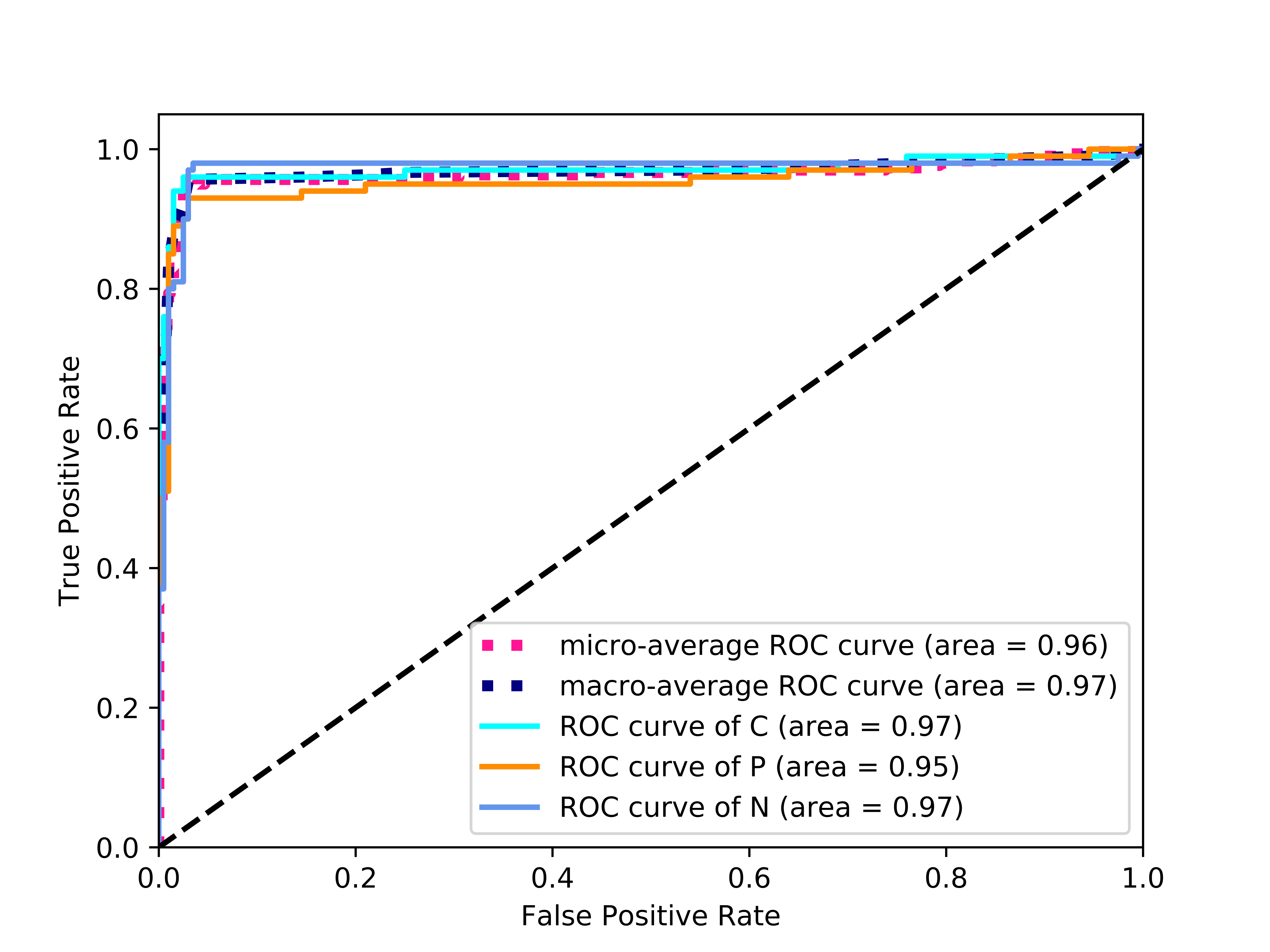} }}
    \caption{Confusion Matrix (a) and ROC curve (b) obtained for Fold-1.}
    \label{fig:ROCCMM}
\end{figure}

The proposed architecture is implemented in Keras \cite{chollet2015keras} with TensorFlow \cite{abadi2016tensorflow} as the backend. The experiments were conducted and evaluated on a 64-bit workstation with a CentOS Linux 7 operating system, hard disk drive, NVIDIA Tesla P100 with 16 GB dedicated GPU, and Intel(R) Xeon(R) Silver 4114 CPU @ 2.20 GHz processor. Datasets and source code are available in the following GitHub repository: \url{https://github.com/ActiveNeuron/MobileCaps}

We evaluated the proposed MobileCaps architecture with 5-fold patient-wise cross-validation. Train, validation, and test data of each fold are prepared with the holdout approach. Table \ref{Table 4} shows the results of 5-fold cross validation and Table \ref{Table 5} shows the performance comparison of the proposed MobileCaps with the benchmark models (all the methods in Table \ref{Table 5} are subjected to 5-fold cross-validation). It is evident from Table \ref{Table 5} that the proposed model outperforms all the benchmarks by a sufficiently large margin in both recall as well as precision. Fig.~\ref{fig:ROCCMM} depicts the confusion matrix and the ROC curve obtained for one of the fold-1. The proposed model achieved a precision of 98.50, which indicates that only 1.50\% of the samples were miss classified into non-COVID-19 Pneumonia and Normal cases (least false positive rate) which is quite significant and a recall rate of 91.60 with an F1-score of 94.91 for COVID-19 cases. The proposed model is able to classify COVID-19 CXR images regardless of the severity level.
Moreover, we compare and evaluate the performance of the proposed model with two more RT-PCR confirmed datasets, namely the Twitter COVID-19 CXR data \cite{Tweet} and JSS COVID-19 CXR data. The results are depicted in Table \ref{Table 7}. We can observe that the proposed model has superior performance than other models by achieving an accuracy of 90.60 and 91.42 on JSS COVID-19 CXR data and Twitter COVID-19 CXR data demonstrating high generalizability. 
\begin{table*}
\renewcommand{\arraystretch}{1.3}
\centering
\caption{Results of 5-Fold Cross Validation   (\textbf{C}: COVID-19 \textbf{N}: Normal \textbf{P}: non-COVID-19 Pneumonia).}
\begin{tabular}{@{}p{3.2cm}cccccccccc@{}}\toprule Models &

\multicolumn{3}{c}{\textbf{PPV/Precision}} &  \multicolumn{3}{c}{\textbf{Sensitivity/Recall}} &
\multicolumn{3}{c}{\textbf{F1 Score}} \\

\cmidrule(lr){2-4} \cmidrule(lr){5-7} \cmidrule(lr){8-10}

& \textbf{C} & \textbf{N} & \textbf{P} & \textbf{C} & \textbf{N} & \textbf{P} & \textbf{C} & \textbf{N} & \textbf{P}  \\ \midrule

Fold1 & 96.88 & 92.08 & 94.17 & 93.00 & 93.00 & 97.00 & 94.90 & 95.24 & 95.37\\

Fold2 & 98.91 & 86.24 & 90.91 & 91.00 & 94.00& 90.00 & 94.79 & 89.95 & 90.45 \\

Fold3 & 98.92 & 83.48 & 95.65 & 92.00 & 96.00 & 88.00 & 95.34 & 89.30 & 91.67\\

Fold4 & 98.89 & 88.79 & 90.29 & 89.00 & 95.00 & 93.00 & 93.68 & 91.79 & 91.63\\

Fold5 & 98.84 & 90.48 & 92.08 & 93.00 & 95.00 & 93.00 & 95.88 & 92.68 & 92.54\\

\textbf{Average} & \textbf{98.50} & 88.21 & \textbf{92.62} & \textbf{91.60} & \textbf{94.60} & \textbf{92.20} & \textbf{94.91} & \textbf{91.22} & \textbf{92.33}\\

\bottomrule
\end{tabular}
\label{Table 4}
\end{table*}

\begin{table*}
\renewcommand{\arraystretch}{1.3}
\centering
\caption{Comparison of the proposed model with other benchmark models (\textbf{C}: COVID-19 \textbf{N}: Normal \textbf{P}: non-COVID-19 Pneumonia).}
\begin{tabular}{@{}p{3.2cm}cccccccccc@{}}\toprule Models &
\multicolumn{3}{c}{\textbf{PPV/Precision}} &  \multicolumn{3}{c}{\textbf{Sensitivity/Recall}} &
\multicolumn{3}{c}{\textbf{F1 Score}} \\

\cmidrule(lr){2-4} \cmidrule(lr){5-7} \cmidrule(lr){8-10}

& \textbf{C} & \textbf{N} & \textbf{P} & \textbf{C} & \textbf{N} & \textbf{P} & \textbf{C} & \textbf{N} & \textbf{P}  \\ \midrule

Eduardo et al.\cite{luz2020towards} & 93.80 & 90.20 & 91.85 & 78.22 & 93.45 & 89.30 & 85.30 & 91.79 & 90.55\\

Afshar et al.\cite{afshar2020covid} & 67.25 & 85.24 & 78.43 & 75.27 & 82.64 & 83.20 & 71.03 & 83.91 & 80.54 \\

Rahimzadeh et al.\cite{rahimzadeh2020modified} & 88.50 & \textbf{90.40} & 90.40 & 87.60 & 92.40 & 90.80 & 88.04 & 91.38 & 90.59\\

\textbf{Proposed Method} & \textbf{98.50} & 88.21 & \textbf{92.62} & \textbf{91.60} & \textbf{94.60} & \textbf{92.20} & \textbf{94.91} & \textbf{91.22} & \textbf{92.33}\\
\bottomrule
\end{tabular}
\label{Table 5}
\end{table*}

\begin{table}
\renewcommand{\arraystretch}{1.3}
\centering
\caption{Comparing the Generalizing Capability of the Proposed Model with the Benchmark Models. }
\begin{tabular}{@{}p{5cm}cc@{}}\toprule Method &
 \multicolumn{2}{c}{\textbf{Accuracy}} \\
\cmidrule(lr){2-3}
& \textbf{JSS COVID-19 CXR Data} & \textbf{Twitter COVID-19 CXR Data} \\
\midrule

Eduardo et al.\cite{luz2020towards}  & 35.76 & 89.31 \\

Afshar et al.\cite{afshar2020covid} & 46.76 & 70.99\\

Rahimzadeh et al.\cite{rahimzadeh2020modified} & 83.58 & 68.42 \\

\textbf{Proposed Method} & \textbf{94.50} & \textbf{91.42} \\

\bottomrule
\end{tabular}
\label{Table 7}
\end{table}

\begin{figure*}[t!]
 \centering
  \includegraphics[scale=0.070]{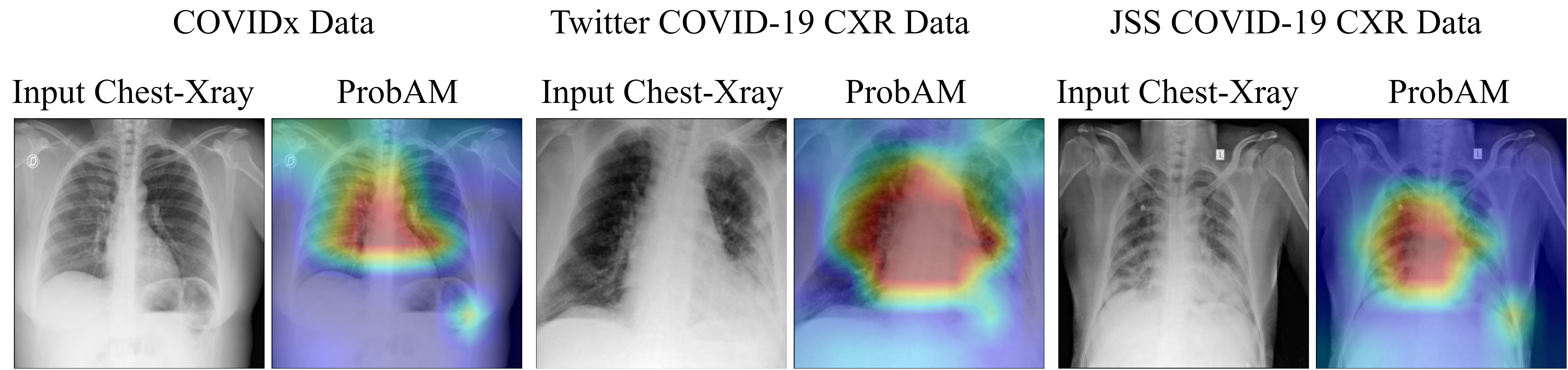}
  \caption{ProbAM \cite{ren2019evaluating} visualization of COVIDx data \cite{lindawangg}, Twitter COVID-19 CXR data \cite{Tweet} and JSS COVID-19 CXR data.}
  \label{fig:Prob}
\end{figure*}

Another notable thing about the proposed architecture is the number of trainable parameters. As can be seen in Table \ref{Table 6}, the proposed model makes use of fewer trainable parameters in comparison to the state-of-the-art models, which helps in reducing the computation cost as well as hardware overhead. Though Afshar et al. \cite{afshar2020covid} have fewer parameters, Its performance on the recent version of COVIDx data was found to be incompetent. Further, In Fig.~\ref{fig:Prob}, we have visualized Probability-guided Activation maps (ProbAM) \cite{ren2019evaluating} of COVIDx data \cite{lindawangg},Twitter COVID-19 CXR data \cite{Tweet} and JSS COVID-19 CXR data as an evidence for the areas focused or learned by the proposed method. ProbAM results clearly show the appropriate targeted ROI region, which the proposed model has learned from the input Chest X-Ray.

\begin{table}[hbt!]
\renewcommand{\arraystretch}{1.3}
\centering
\caption{Parameter Comparison.}
\begin{tabular}{p{3.5cm}cr}\toprule
\textbf{Model} &\phantom{abc}& \textbf{Parameters}  \\
\midrule
Rahimzadeh et al.\cite{rahimzadeh2020modified} &&  48M \\
Karim et al. \cite{karim2020deepcovidexplainer} && 21M\\
Eduardo et al.\cite{luz2020towards}         && 11.60M \\
Oh et al.\cite{oh2020deep}                 && 11.60M\\
Wang et al.\cite{wang2020covid}             && 11.75M \\
Khobahi et al.\cite{khobahi2020coronet}    && 11.80M\\
Ucar et al.\cite{ucar2020covidiagnosis} && 724K \\
Afshar et al.\cite{afshar2020covid}         && 295K \\

\textbf{Proposed Method}   && 
\textbf{2.2M}\\

\bottomrule
\label{Table 6}
\end{tabular}
\end{table}

Due to the unavailability of the source code and insufficient information; we were not able to perform a one-to-one comparison with some of the methods, but based on the number of samples on which the proposed model is evaluated, the number of trainable parameters, and the generalizing ability we strongly argue that our model outperforms the recent approaches. In  \cite{oh2020deep} the authors followed a patch-wise approach based on the statistical interpretation of the data. Authors compared the performance of their model with COVID-Net \cite{wang2020covid}, and as per their reported results, their approach managed to reach a sensitivity of 100\% and a precision of 91.90\% on COVID-19 cases. One noteworthy fact is that the authors evaluated the performance of their model on test data of 300 samples with just 10 samples of COVID-19 cases. COVID-Net \cite{wang2020covid} is considered as the state-of-the-art for detecting COVID-19 CXR. Authors of COVID-Net \cite{wang2020covid} achieved a precision and recall of 98.9\% and 91.0\% respectively for COVID-19 cases; whereas, the proposed method achieved a precision of 98.50, which is comparable with 98.9 of \cite{wang2020covid} and a recall of 91.60 after performing 5-fold cross-validation. Moreover, the proposed model is proven to be highly generalizable from Table \ref{Table 7}. Further, our model uses 81.27\% fewer parameters than COVID-Net \cite{wang2020covid}. The performance analysis of the proposed MobileCaps-S is also evaluated with a 5-fold cross-validation technique and achieved a mean R$^2$ score of 70.51 $\pm$ 2.3504 in predicting the RALE severity of COVID-19 CXR. The model outputs a severity score in the range of 1-8, which can be used to determine the level of infection.

\section{Conclusion}
In this study, we presented MobileCaps, a fully automated, hybrid, and lightweight model for the accurate classification of COVID-19 CXR images, as well as MobileCaps-S for predicting severity scores based on the RALE scoring technique. The proposed models showed an effective way to utilize a Convolutional Neural Network to form powerful primary capsules in a Capsule Network that could construct a rich feature space with a great reduction in the number of trainable parameters. The MobileCaps model was trained and extensively evaluated on the COVIDx dataset, and further, the generalizing ability of the model was evaluated with two more additional datasets. Saliency maps visualized using the ProbAM technique depicted the appropriate ROI region learned by the proposed model, which further solidified its generalizability. Additionally, the COVID-19 CXR images from the COVIDx dataset were annotated using the RALE scoring technique and trained with the MobileCaps-S model for severity assessment of COVID-19. The superior performance of the proposed models, as corroborated by the experimental results, coupled with their lightweight nature, makes them an ideal candidate for easing the burden of healthcare systems compared to other state-of-the-art models in the literature.
\bibliographystyle{unsrt}
\bibliography{ref}
\end{document}